\documentclass[14pt,aps,prl,twocolumn,nofootinbib,showpacs,superscriptaddress,groupedaddress]{revtex4-1}
\usepackage{graphicx,epsfig}
\usepackage{bm}
\usepackage{subfigure}
\usepackage{natbib}
\begin{document}

\title{The role of structure and entropy in determining differences in dynamics for glass formers with different interaction potentials}

\author{Atreyee Banerjee$^1$,  Shiladitya Sengupta$^2$,  Srikanth Sastry$^{2,3,\dagger}$,  Sarika Maitra Bhattacharyya$^{1}$}
\email{mb.sarika@ncl.res.in,$^\dagger$ sastry@tifrh.res.in}
\affiliation{$^1$ Polymer Science and Engineering Division, CSIR-National Chemical Laboratory, Pune-411008, India}

%\author{Shiladitya Sengupta}
\affiliation{$^2$TIFR Centre for Interdisciplinary Sciences, 21 Brundavan Colony, Narsingi, Hyderabad 500075, India}
%\author{Srikanth Sastry}
\affiliation{$^3$Theoretical Sciences Unit, Jawaharlal Nehru Centre for Advanced Scientific Research, Jakkur Campus, Bangalore 560 064, India}
\affiliation{TIFR Centre for Interdisciplinary Sciences, 21 Brundavan Colony, Narsingi, Hyderabad 500075, India}
%\author{Sarika Maitra Bhattacharyya}

%\affiliation{Department of Polymer Science and Engineering, CSIR-National Chemical Laboratory, Pune-411008, India}

%\begin{document}

%\date{\today}

\begin{abstract}
We present a study of two model liquids with different interaction
potentials, exhibiting similar structure but significantly different
dynamics at low temperatures. By evaluating the configurational
entropy, we show that the differences in the dynamics of these systems
can be understood in terms of their thermodynamic
differences. Analyzing their structure, we demonstrate that
differences in pair correlation functions between the two systems, through
their contribution to the entropy, dominate the differences in their
dynamics, and indeed overestimate the differences. Including the
contribution of higher order structural correlations to the entropy
leads to smaller estimates for the relaxation times, as well as smaller
differences between the two studied systems.
\end{abstract}

%\keywords{supercooled liquid, fragility, glass transition}
%\pacs{ 64.70Q-, 64.70pm, 61.20LC}
\maketitle 

Many approaches towards understanding the dynamical behavior of liquids attempt to predict dynamics in terms of static structural 
correlations \cite{Gotzebook,andersen2005}, often focussing on two-body correlation functions. In turn, it has been argued that the short range, 
repulsive interactions have a dominant role in determining the pair correlation function, with the attractions making a perturbative contribution. 
Such an approach was shown to be effective in predicting the pair correlation function for dense liquids interacting {\it via.} the Lennard-Jones (LJ) potential, by Weeks,
 Chandler and Andersen, who treated the LJ potential as a sum of a repulsive part (referred to subsequently as the WCA potential) and the attractive part \cite{chandler}. 
If such a treatment carries over to the analysis of dynamics, the expectation would be that liquids with LJ and the corresponding WCA interactions 
should have similar dynamics. However, in a series of recent papers, Bertheir and  Tarjus have shown that model liquids with LJ and WCA interactions, 
exhibiting fairly similar structure, exhibit dramatically different dynamics, characterized by a structural relaxation time, at low temperatures 
\cite{tarjus_prl,tarjus_pre,tarjus_epje,tarjus_berthier_jcp}. In order to analyze this ``non-perturbative'' effect of the attractive forces on the dynamics,
 Berthier and Tarjus studied a number of ``microscopic'' approaches to predict the dynamics, based on knowledge of the static pair correlations. They conclude that
 the approaches they analyze are unsuccessful in capturing the differences in dynamics between the LJ and WCA systems. Dyre and co-workers \cite{Toxvaerd, Pedersen-prl,Bohling} 
have argued that the origins of these observations are not specifically in the inclusion or neglect of attractive interactions\cite{Bohling},
 but factors such as the inclusion of interactions of all first shell neighbors \cite{Toxvaerd}, and the presence or absence of scaling between systems/state points 
compared \cite{Pedersen-prl}. In particular, Pedersen and Dyre \cite{Pedersen-prl} identify a purely repulsive inverse-power-law (IPL) potential that has dynamics that
 can be mapped to the LJ case studied by Bertheir and Tarjus.  These observations notwithstanding, the inability to capture the differences between the LJ and WCA system 
highlighted by Berthier and Tarjus by predictive approaches to dynamics remains an open issue. In this regard, it has been suggested by Coslovich \cite{coslovich,coslovich-jcp} 
that higher 
order structural correlations may play a significant role in determining dynamics, and he argues this point by showing that the temperature variation of locally preferred 
structures for the LJ and WCA systems tracks that of the relaxation times \cite{coslovich}. Hocky {\it et al.} \cite{hocky-prl} show, by evaluating the {\it point-to-set} length 
scales 
in the LJ, WCA and IPL liquids, that while the LJ and IPL liquids show essentially the same temperature dependence, the WCA system differs from these two, 
thereby offering a quantitative explanation of the dynamics, in terms of a quantity that has implicit dependence on two body and many body structural correlations. 

Among the prominent predictive relationships between equilibrium properties and dynamics for liquids at low temperatures is the Adam-Gibbs relation \cite{adam-gibbs}, 
\begin{equation}
\tau(T)=\tau_{o}\exp\left(\frac{A}{TS_{c}}\right),
\label{ag}
\end{equation}
which expresses relaxation times $\tau$ in terms of a thermodynamic quantity, the configurational entropy $S_c$.
 The usefulness of  this relationship, whose rationalization has a close relation theoretically with the growing static length scales explored by Hocky 
{\it et al.} \cite{hocky-prl}, in comprehending the differences in dynamics of the LJ and WCA systems has not been hitherto explored. 

In this Letter, we test whether the differences in the interaction potential between the LJ and WCA systems, while having a modest effect on structure, have 
a more significant effect on the thermodynamics, and the 
Adam-Gibbs (AG) relation can hence capture the quantitative differences in the dynamics between these systems. We further employ this relation as a tool
 to explore the contributions of two body and higher order structural correlations, by considering a two body approximation to the configurational entropy. 
We find that: (1) The Adam-Gibbs relationship quantitatively captures the differences in the dynamics between the LJ and WCA systems. (2) Two body correlations alone,
 used to obtain an approximation to the configurational entropy, yield a significant difference in predicted relaxation times, indeed overestimating the difference, 
indicating a strong sensitivity to changes in pair correlations. Reminiscent of the predictions from mode coupling theory (MCT) calculations, however, 
the relaxation times are significantly overestimated using only the two body approximation to the entropy. (3) The residual multiparticle entropy (RMPE), 
arising from many particle correlations, speeds up the dynamics at low temperatures and is larger for LJ system,  which is at odds with the notion that stronger 
multiparticle correlations are responsible for the stronger temperature dependence of the relaxation times but consistent with the observation that a significant
 contribution to higher order (three body) correlations arise from the amplification of small differences in correlation at the two body level \cite{coslovich,coslovich-jcp}. 

\begin{figure}[h]
\centering
\subfigure{
\includegraphics[width=0.45\textwidth]{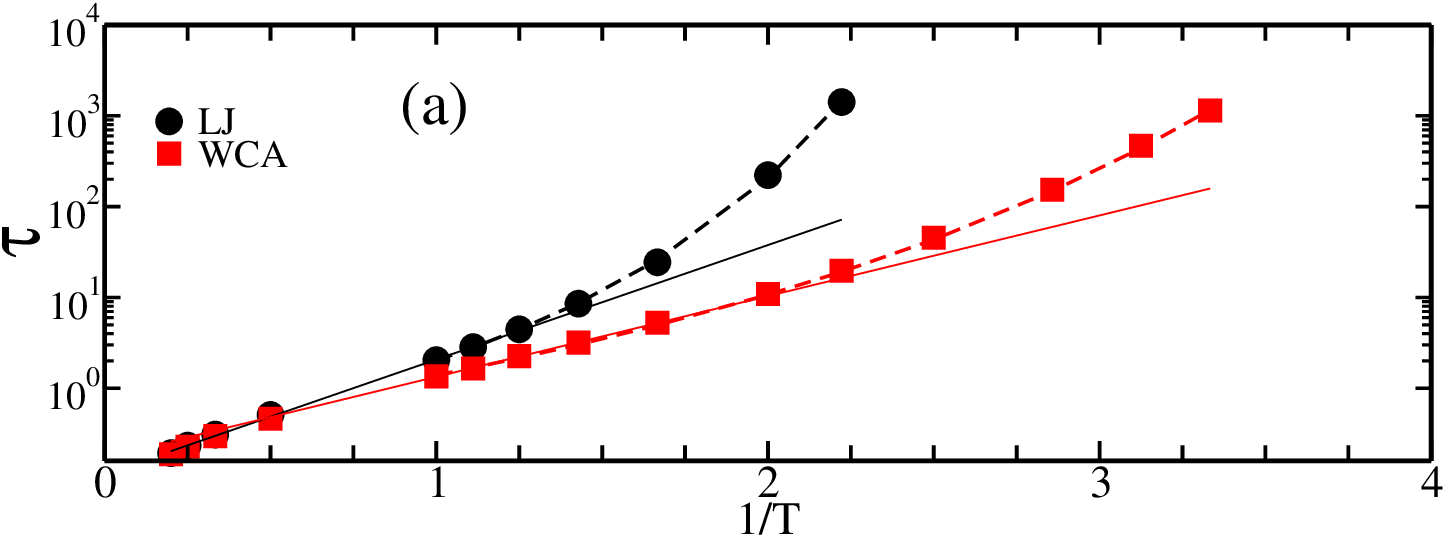}}
\label{VFT}%

\subfigure{
\includegraphics[width=0.45\textwidth]{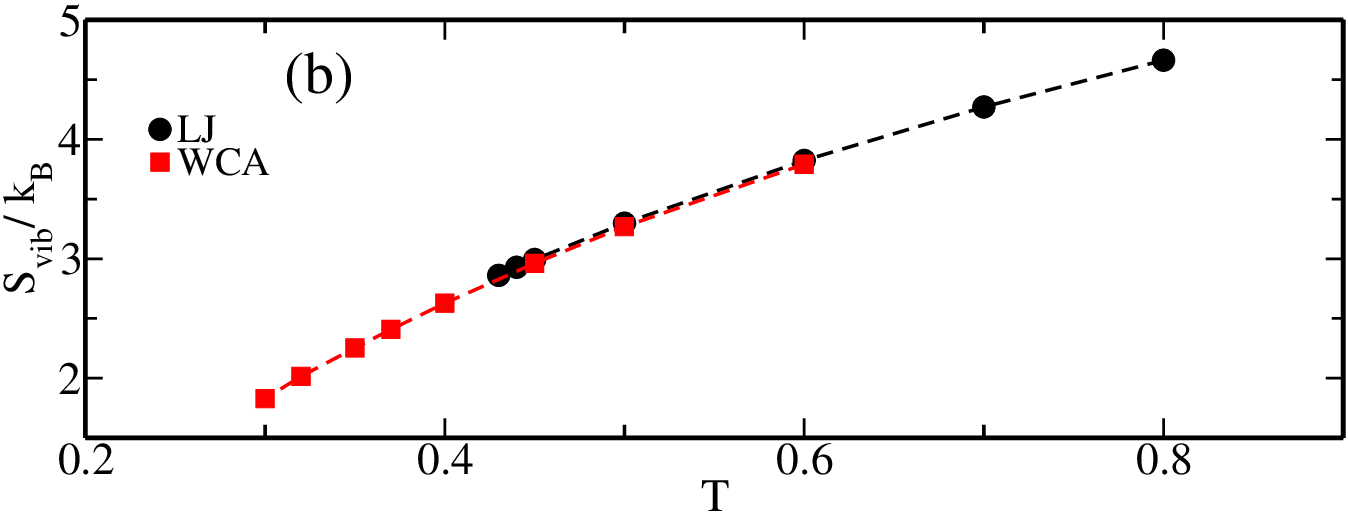}}

\subfigure{
\includegraphics[width=0.45\textwidth]{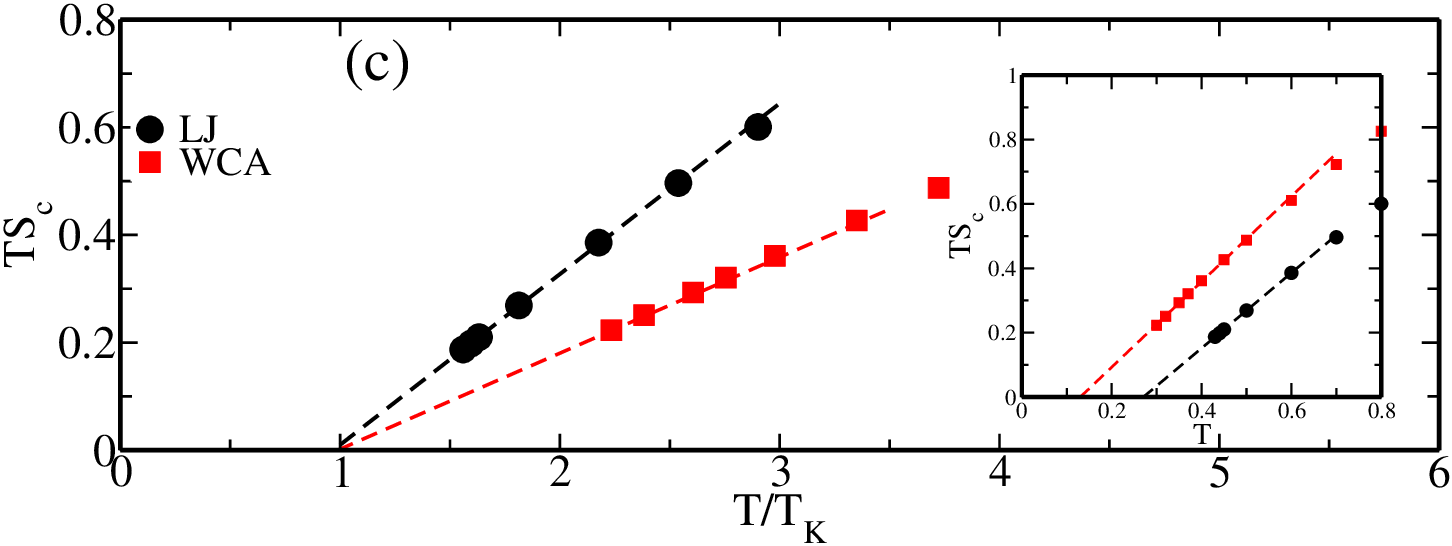}}

\subfigure{
\includegraphics[width=0.45\textwidth]{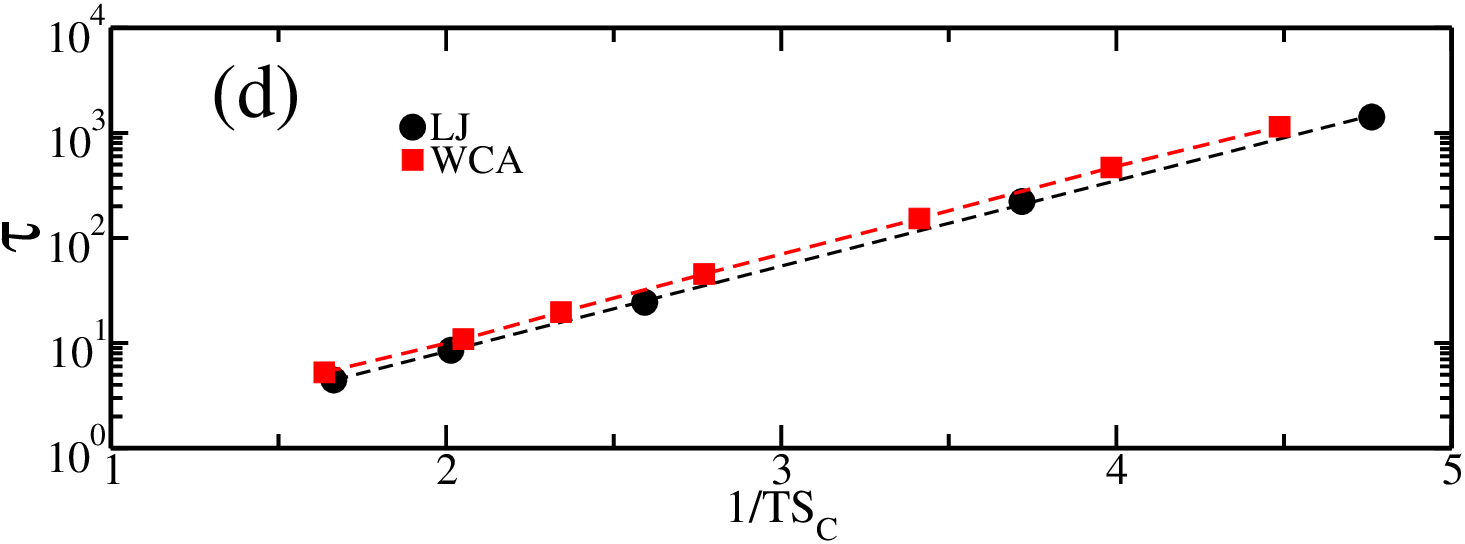}}

\caption{(a) Arrhenius plot of the relaxation time $\tau$ defined from overlap function  $q( t=\tau)= 1/e$. High temperature Arrhenius fits and low
 temperature VFT fits are also shown.(b)The temperature dependence of vibrational entropy ($S_{vib}$)  for LJ and WCA systems.
(c) Determination of thermodynamic fragility, $K_T$ from the slope of the linear fit. (inset) $T_K$ is the Kauzmann temperature 
obtained from the linear fit from $S_c(T_K)=0$. $T_K$ is 0.27 and 0.134 for LJ and WCA systems respectively. (d) The Adam-Gibbs plot, showing that the differing 
temperature dependence of relaxation times are quantitatively captured by the temperature variation of the configurational entropy.}
\label{fig1}
\end{figure}

We study the LJ and WCA versions of the Kob-Andersen binary mixture at $\rho=1.2$, where the difference in dynamics between
 the two systems are pronounced \cite{tarjus_berthier_jcp} with simulation details as in \cite{Srikanth_nature,tarjus_berthier_jcp}. 
Lengths, temperatures, and times are given in units of $\sigma_{AA}$, $\epsilon_{AA}/k_B$, $(m\sigma_{AA}^2/\epsilon_{AA})^{1/2}$ respectively.
 We calculate the relaxation time $\tau$ from the overlap function $q(t)$ as described in reference \cite{shila-jcp}, by the condition $q(t=\tau)= 1/e$. 
The temperature dependence of the relaxation times shown in an Arrhenius plot in Fig.\ref{fig1} (a) illustrate, as discussed earlier \cite{tarjus_prl}, 
that the LJ system has a much stronger temperature dependence than the WCA system. We quantify the temperature dependence by fitting $\tau (T)$ to the Vogel-Fulcher-Tammann 
(VFT)
 expression, $ \tau(T)=\tau_{o}\left[ \frac{1}{K_{VFT}(\frac{T}{T_{VFT}}-1)}\right]$. The resulting {\it kinetic} fragilities for the two systems are $K_{VFT} = 0.19$ 
for the LJ liquid and $0.14$ for WCA, with divergence temperatures $T_{VFT} = 0.28$ and $0.16$ respectively, with the ratio $K_{VFT}^{LJ}/K_{VFT}^{WCA} = 1.36$. 
The VFT form can be obtained from the AG relation if $TS_{c}=K_{T} \left(\frac{T}{T_{K}}-1\right)$, 
with the kinetic fragility $K_{VFT}$ given in terms of the {\it thermodynamic} fragility $K_{T}$ (with $T_{K} = T_{VFT}$) by $K_{VFT} = K_{T}/A$. 
The configurational entropies (per particle) are calculated as the difference between the total and vibrational entropies, $S_{c}=S_{total}-S_{vib}$.\footnote{as 
described  in {\it e. g.} \cite{Srikanth_nature} where the Planck's constant is calculated using Argon unit} As shown in Fig.\ref{fig1}(b) the vibrational entropies are
 similar for the two systems.  
In Fig.\ref{fig1} (c) (inset)  we show that by extrapolation, $S_c$ for the LJ system vanishes at a higher
 temperature, and (main panel) has a higher thermodynamic fragility $K_{T}$. Fig.\ref{fig1} (d) shows the Adam-Gibbs plot, $\tau$ {\it vs.} $1/TS_c$. For both the LJ
 and WCA systems, the AG relation is not only valid, but the slopes $A$ for the two systems (related to the high temperature activation energy $E_{\infty}$ and the 
limiting value of $S_c$, $S_c^{\infty}$ by $A = E_{\infty} S_c^{\infty}$) are very close. Thus, the temperature variation of the configurational entropy $S_c$ fully 
captures the differences in the dynamics between these two systems.

\begin{figure}[h]
\centering
\begin{subfigure}
\centering
\includegraphics[width=0.45\textwidth]{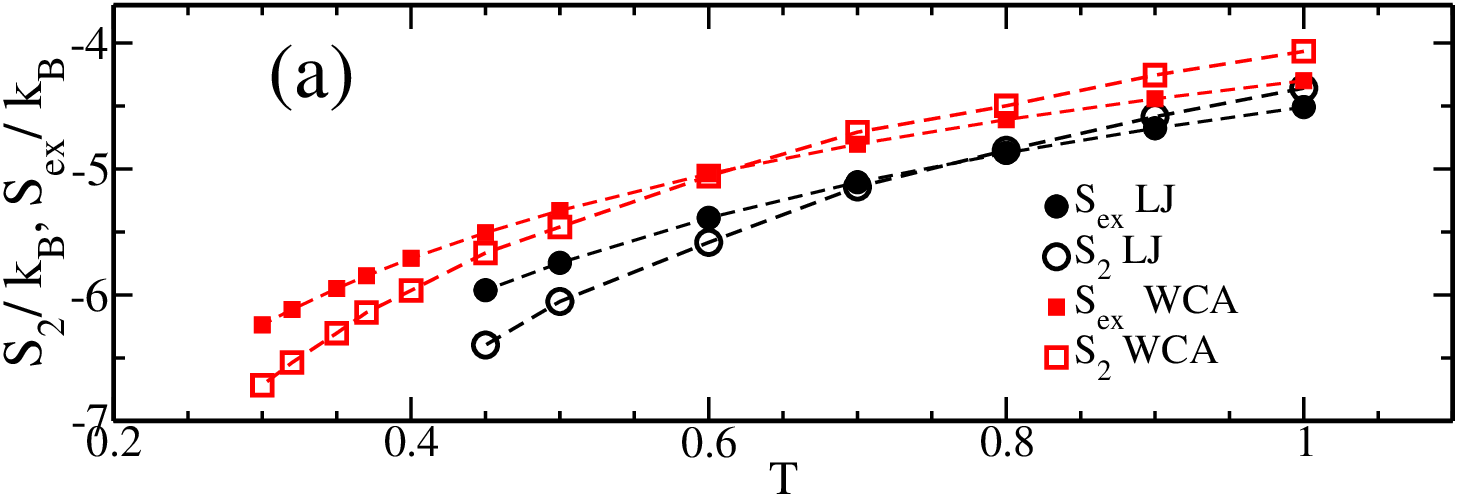}
\label{crossover}
\end{subfigure}

\begin{subfigure}
\centering
\includegraphics[width=0.45\textwidth]{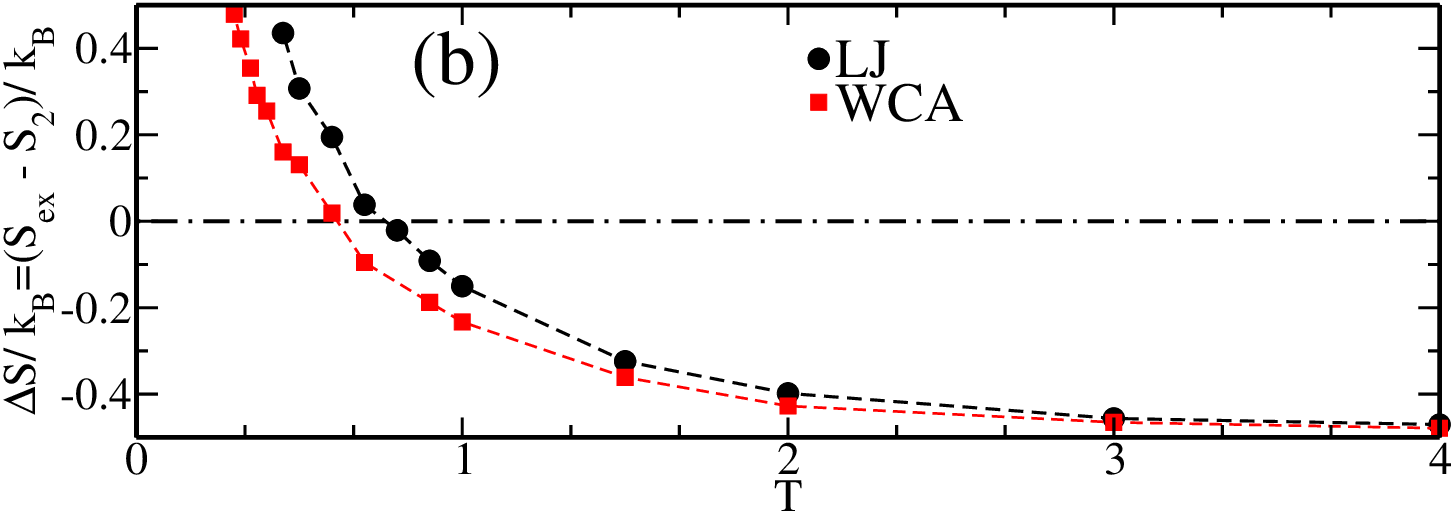}
\label{deltas}
\end{subfigure}
\caption{(a) Plots of $S_{ex}$ and $S_{2}$ {\it vs.} temperature, showing that the two quantities cross at intermediate temperatures for both models.The 
crossover temperatures are 0.77 and 0.61 for  LJ and  WCA systems respectively. (b) The residual multiparticle entropy, $\Delta S=(S_{ex}- S_2)$ vs. temperature. 
The connecting lines are guide to the eye.}
\label{fig2}
\end{figure}

In order to discuss the contribution of two body and higher order static correlations to the dynamics, we consider the per particle excess entropy $S_{ex}$,
 defined by $S_{total} = S_{id}+S_{ex}$ where $S_{id}$ is the ideal gas entropy (per particle) . $S_{ex}$ can be expanded in an infinite series, 
$S_{ex}=S_{2}+S_{3}+.....=S_{2}+\Delta S$ using Kirkwood's factorization \cite{Kirkwood} of the N-particle distribution function \cite{green_jcp,raveche,Wallace}. 
 $S_{n}$ is the $``n"$ body contribution to the entropy. Thus the pair excess entropy is $S_{2}$ and the higher order contributions to excess entropy is given by 
the residual multiparticle entropy (RMPE), $\Delta S=S_{ex}-S_{2}$
 \cite{giaquinta-1,*giaquinta-2}. $S_{2}$ for a binary system can be written in terms of 
the partial radial distribution functions,
\begin{equation}
\frac{S_{2}}{k_{B}}=-\frac{\rho}{2} \sum_{\alpha,\beta}x_{\alpha} x_{\beta} \int_0^{\infty} d {\bf r}\{g_{{\alpha}{ \beta}}(r) \ln g_{{\alpha} {\beta}}(r)- [g_{{\alpha}{\beta}}(r)-1]\}
\label{s2}
\end{equation}
\noindent where $ g_{{\alpha}{ \beta}}(r)$ is the atom-atom pair correlation between atoms of type $\alpha$ and $\beta$, $N$ is the total number of particles,
 $x_{\alpha}$ is the mole fraction of component $\alpha$ in the mixture, and $k_B$ is the Boltzmann constant. In Fig.\ref{fig2}(a) we show a comparison of $S_{ex}$ 
and $S_2$, and in Fig.\ref{fig2}(b) we show the RMPE, $\Delta S$. Interestingly, for both the LJ and WCA systems, starting out at high temperatures being larger than $S_{ex}$ 
as one may expect, $S_2$ becomes smaller than $S_{ex}$ at low temperatures. This behavior, previously noted on other contexts \cite{baranyai_PRA,truskett_2008, Murari_charu_jcp},
 means that the RMPE, arising from many body effects, is positive at low temperatures. This change in sign in RMPE implies that although many body correlations at high temperature
slows down the dynamics as may be expected, at low temperature their role is reversed.
 Further, we note that the RMPE is at all temperatures bigger for LJ than WCA - thus the role of many body correlation at low temperatures is to increase the entropy 
 and to a greater extent for the LJ than WCA system.

\begin{figure}[h]
\centering
\includegraphics[width=0.47\textwidth]{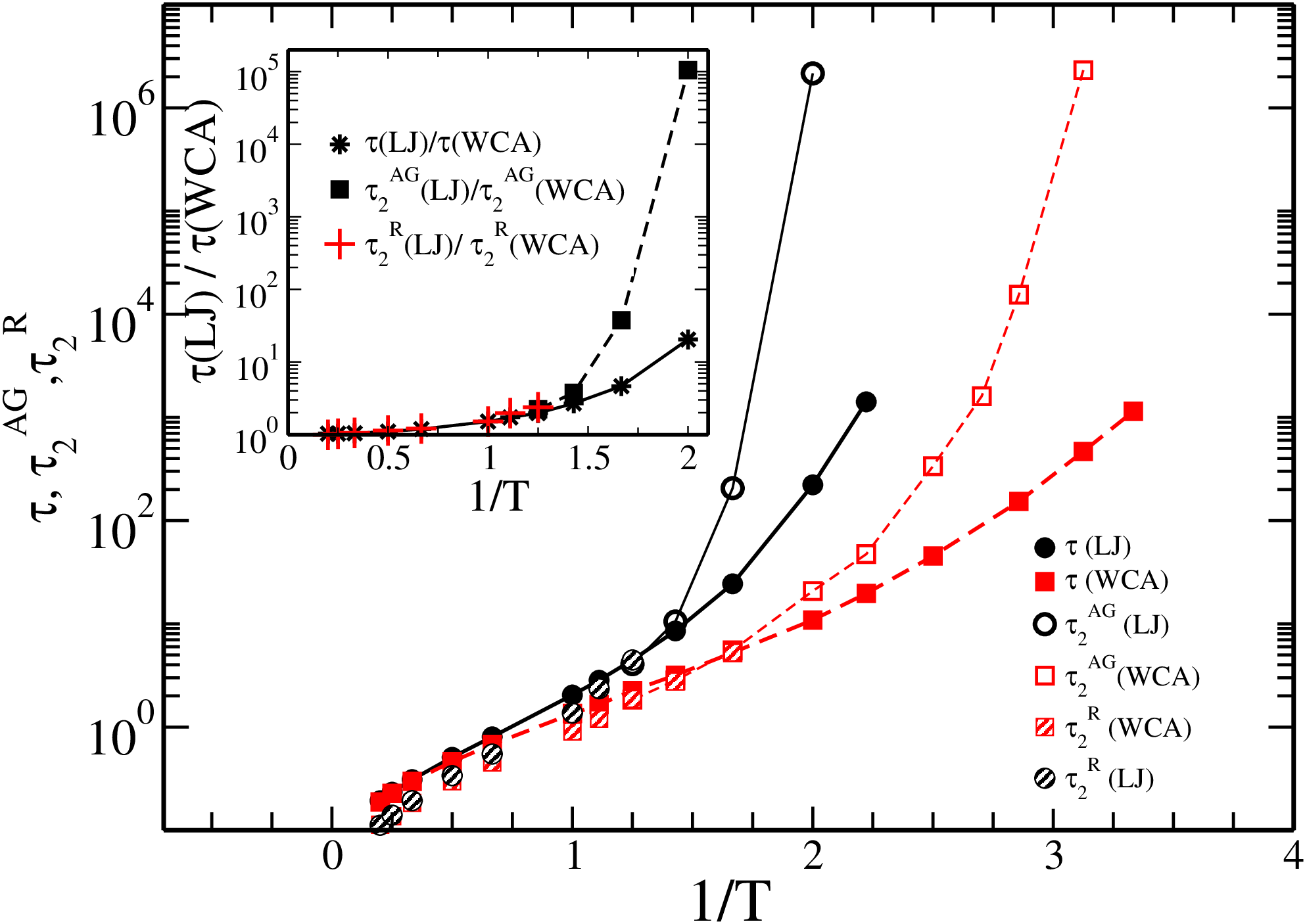}
%\label{adam-gibbs}
\caption{ $\tau$, $\tau_{2}^{AG}$, $\tau_2^{R}$ vs. $1/T$ for LJ and WCA systems. (inset) Their ratios for LJ and WCA systems vs. $1/T$. }
\label{fig3}
\end{figure}

At normal liquid temperatures, a semi quantitative correlation between the dynamics (transport properties) 
and thermodynamics (excess entropy), proposed by Rosenfeld \cite {Rosenfeld-iop,rosenfeld}, has been extensively studied in recent times,
 with the form $\tau(T)=C \exp \left[ -KS_{ex}\right]$ where $C$ and  $K$ are constants. Since the pair entropy $S_{2}$ accounts for $80\%-90\%$ of the excess entropy
 \cite {Baranyai-cp} (Fig.\ref{fig2} (a)), many studies replace $S_{ex}$ by $S_{2}$ \cite{tarjus_berthier_jcp,Dzugutov,hoyt, trusket_2006,Ruchi_charu_2006}.
For the systems studied here this approximation is found to hold good for high T. We can write 
\begin{equation}
\tau(T)=C \exp \left[ -KS_{ex}\right]=\tau^{R}_{2}(T)* \exp \left[ -K\Delta S\right]
\end{equation}
\noindent
where $\tau^{R}_{2}(T) =C \exp \left[ -KS_{2}\right]$. The $C$ and $K$ are obtained from linear fits of $\ln \tau(T)$ against $S_{ex}$ at high temperatures 
(above the temperatures 
T=0.8 and 0.6 for LJ and WCA respectively).
 The $\tau^{R}_{2}$ thus obtained, plotted for the LJ and WCA systems in Fig.\ref{fig3} for high to intermediate temperatures, 
agree well with $\tau(T)$ since the contribution from $\Delta S$ is only about $ 10\%$ of $S_{ex}$. As shown in the inset of Fig.\ref{fig3}, 
the ratio of $\tau$ values for LJ and WCA are well approximated by that obtained with $\tau^{R}_{2}$. 

We next turn to the role of two body and higher order correlations in determining the dynamics as reflected in the configurational entropy. 
 To get an estimate of the configurational entropy as predicted by the pair correlation we rewrite $S_{C}$ in terms of the pair contribution to 
configurational entropy $S_{C2}$,
\begin{equation}
S_{C}=S_{id}+S_{ex}-S_{vib}=S_{id}+S_{2}+\Delta S-S_{vib}=S_{C2}+\Delta S
\end{equation}
\noindent
Where $S_{C2}=S_{id}+S_{2}-S_{vib}$.  As mentioned earlier the vibrational entropies of the LJ and WCA systems are found to be very close to each other(Fig.\ref{fig1} (b)). 
However, 
the apparently similar structures predict different $S_{2}$  (Fig.\ref{fig2}a) and $S_{C2}$ values. We obtain the thermodynamic fragilities, $K_{T2}$ 
as predicted by $S_{C2}$ following the same procedure as described for $S_{C}$ and find the LJ system to be more fragile. 
{\it Thus even considering only two body correlations we find the LJ and the WCA systems to be thermodynamically different}. This finding is similar to the observation 
\cite{coslovich-jcp} that significant changes in thermodynamic properties and also higher order correlation functions may arise as a result of amplification of small 
changes in the pair correlations. 

To determine the effect of pair correlations on the low temperature dynamics we estimate the relaxation times as predicted by accounting only
 for two body correlations, $\tau^{AG}_{2}$. 
To this end, we re-express the AG relation as follows: 
\begin{eqnarray}
\tau(T)&=&\tau_{o}\exp\left(\frac{A}{TS_{c}}\right)=\tau^{AG}_{2} (T)*\exp\left(-\frac{A*\Delta S }{TS_{C2}S_{C}}\right) 
\label{equ_tau_estm}
\end{eqnarray}
\noindent
where $\tau^{AG}_{2}(T) =\tau_{o}\exp\left(\frac {A}{TS_{C2}}\right)$. The $\tau^{AG}_{2}$ for the LJ and the WCA systems are plotted in Fig.\ref{fig3},
 as well as their ratio (inset). We find that $\tau^{AG}_{2}$s diverge at higher temperatures and their values for the LJ and WCA systems are widely different,
 reminiscent of the behavior of relaxation times according to MCT calculations. As seen in the inset of Fig.\ref{fig3}, the ratio of relaxation times is overestimated 
by the corresponding ratio of $\tau^{AG}_{2}$. The kinetic fragility  $K_{VFT2}$ as obtained by fitting the temperature dependence of  $\tau^{AG}_{2}$ to a VFT form, 
shows that the LJ system is more fragile. Their ratio, $\frac{K_{VFT2}^{LJ}} {K_{VFT2}^{WCA}}=1.94$, is bigger than $1.36$ obtained from $\tau(T)$. Thus, considering only the
 two body contribution to the entropy, the Adam-Gibbs relation over estimates the difference in the dynamics between the LJ and WCA systems, rather than fail to capture 
differences between them contradicting the expectation that the pair correlation contributions yield similar dynamics, and that the many body correlations may drive the
 difference between the two systems. Instead, the role of many body correlations, other than lowering the predicted relaxation times for both systems, is also to reduce 
the predicted difference between them. We also note that although the value of $\Delta S$ is similar over the whole temperature regime, it plays a greater role
 at low temperatures.

\begin{figure}[h]
\centering
%\subfigure{
%\includegraphics[width=0.4\textwidth]{fig1a.eps}}
%\label{kintetic_fragility}
%
%\subfigure{
\includegraphics[width=0.4\textwidth]{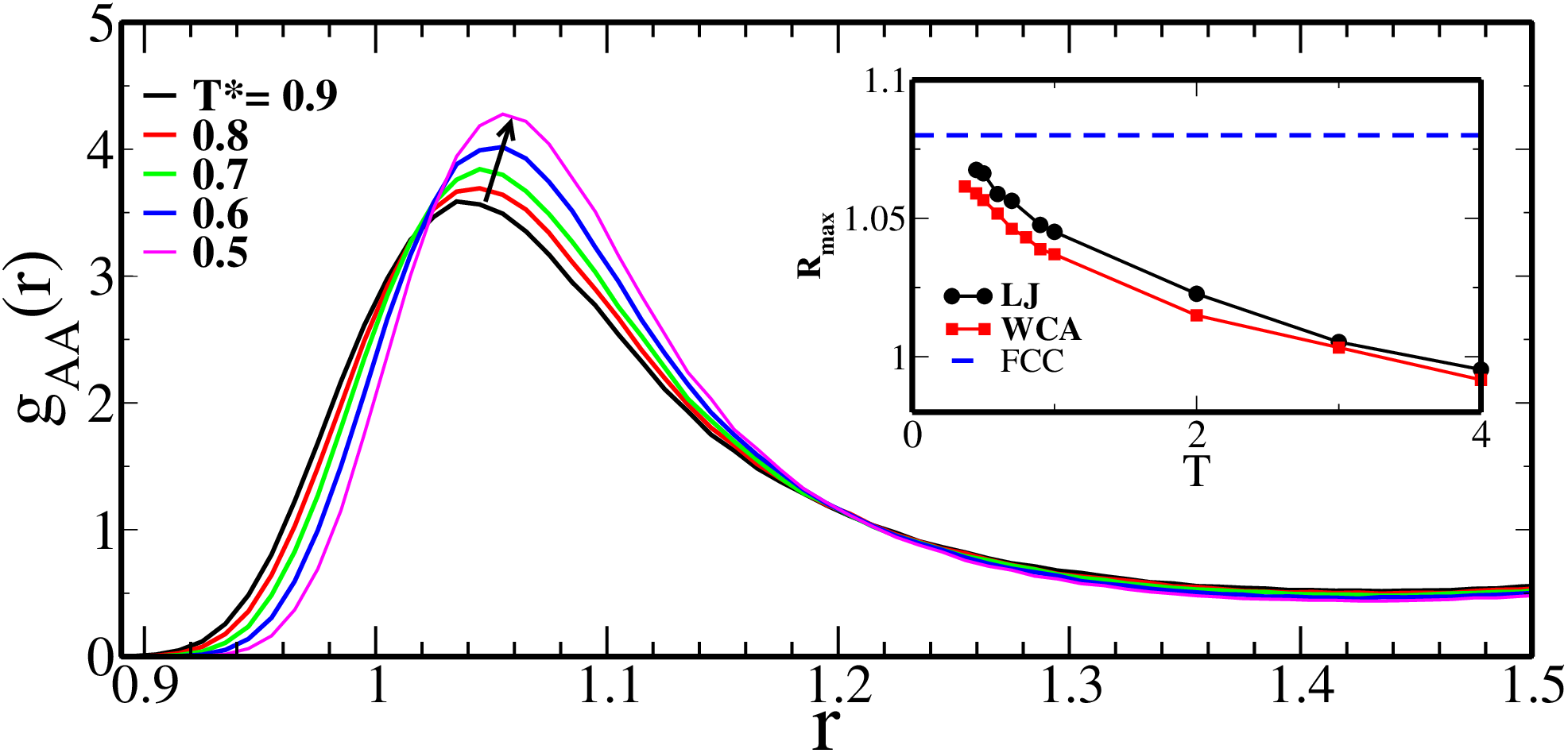}
%\label{adam-gibbs}
\caption{ The first peak position, $R_{max}$ of the radial distribution function shifts to the right with decreasing temperature. 
(inset) $R_{max}$ moves towards the position of the fcc lattice.}
\label{fig4}
\end{figure}

The increase in  $\Delta S$ with decreasing $T$  is usually associated with some ordering in the system 
 \cite{giaquinta-1,*giaquinta-2, trusket_2006,murari_charu_2010_jcp} which we now show is also reflected in the pair correlation function.
 The  first peak of the pair correlation function shifts to the right \cite{cao} as temperature decreases, as shown in Fig. \ref{fig4}. Earlier studies have shown that the 
``A" particles in both the  LJ and WCA
 models show a tendency towards fcc ordering \cite{coslovich,dyre,atreyee}. As shown in the inset Fig. \ref{fig4}, the first peak position of the pair correlation function 
indeed moves towards the value for the {\it fcc} lattice as temperature is lowered. 

In summary, we have shown that the temperature dependence of the configurational entropy, {\it via.} the Adam-Gibbs relation, explains quantitatively the differences in the
 dynamics between the LJ and WCA systems we study. Using an approximation, only two body correlation information, to the configurational entropy,
 we have shown that these correlations capture the differences in the dynamics between the two systems, indeed overestimating the differences, contrary to 
the expectation that the similarity of pair correlation functions between the two systems lead to similar predictions for the dynamics. The contributions from the many body 
correlations speed up of the dynamics thus significantly correcting the overestimation of the relaxation times as solely predicted
 by the pair correlation information and this effect is found to be more for the LJ system.

%\footnotetext[17]{$^{\*}$mb.sarika@ncl.res.in}
%\footnotemark{sastry@tifrh.res.in}
%\clearpage

%\clearpage

%

%\bibliographystyle{unsrt}

%\bibliography{BSSB_v5}
%\footnotemark{mb.sarika@ncl.res.in,sastry@tifrh.res.in}
%\footnote{\star} mb.sarika@ncl.res.in
%\begin{thebibliography}{12}
 
%\end{thebibliography}
\clearpage
%\section{Supporting material: Determination of the thermodynamic fragility}
%\begin{figure}[h]
%\centering

%\includegraphics[height=3cm,width=5cm]{fig2b.eps}
%\label{kintetic_fragility_estm}
%
%\caption{  Determination of thermodynamic fragility,$K_{T2}$, from the slope of the $TS_{C2}$ vs. $T/T_{K2}$ plot using the relation %$TS_{c2}=K_{T2}((T/T_{K2})-1)$ for LJ and WCA system (Table I). The estimated Kauzmann temperature, $T_{K2}$ are 0.43, 0.27 for LJ and WCA %system respectively.  }
%\label{supp_estimated}
%\end{figure}

\end{document}